# Monitoring Systems and Services

Alwin Brokmann
*DESY Hamburg, Germany*

The DESY Computer Center is the home of O(1000) computers supplying a wide range of different services Monitoring such a large installation is a challenge. After a long time running a SNMP based commercial Network Management System, the evaluation of a new System was started. There are a lot of different commercial and freeware products on the market, but none of them fully satisfied all our requirements. After re-valuating our original requirements we selected NAGIOS as our monitoring and alarming tool. After a successful test we are in production since autumn 2002 and are extending the service to fully support a distributed monitoring and alarming.

## 1. INTRODUCTION

Monitoring large installations of different computer systems with lots of applications is a very difficult task

One of the main Problems in Computer Centers is, to find out Problems before Users will run into them.
.
Efficient monitoring will observe the system activities, check if applications are running and will do specified alarming. Service checking can be done in two different way's, monitoring processes or real Service Monitoring. What does it mean if your HTTP-Process is running, is it possible to connect to your Pages? Connecting to a specific Port/ reading the given answer makes real service checking true.

Also a Monitoring-Tool should supply useful informations for System-Administration.

## 2. SOME REASONS FOR US TO SELECT NAGIOS

First off all NAGIOS fulfills the most of our requirements. It is an open System and so it is possible to submit our own tests (Plug in Concept). We need scalability to do Load Sharing on the Monitoring Hosts.

NAGIOS comes with a very simple WEB-Interface, which makes it easy to use. A lot of out of the Box functionality is also included. NAGIOS is able to do some Reporting and Alarming.

Stability of NAGIOS is very excellent, no instabilities or crashes happened in more than two years of using it, of course there are still some bugs.

### 2.1. DESY Computer Center Monitoring Setup

At present we have five Linux Boxes doing all the Monitoring. Three of them are doing Host and Service Checks. One PC is our Central Log-Host, running LogSurfer, a Tool for SYSLOG analysis. The last PC is our Central-Monitoring Host which handles the WEB-. Interface.

Service Data transmission between Hosts and the Monitoring System is done by an PERL Daemon.

The Distributed Monitoring Server ask the Daemon for the specified Information. If the answer is out of given specifications, or if an timeout appears an alarm is triggered. For some reasons this is not done for all our Hosts, we have also passive Service Checks running. If possible we use Service Checking- connecting to an Port or asking a Process for Status details. If it's impossible we use Process Checking instead.

Check-Programs are executed on different Hosts and the results are transferred to the Central Monitoring Host. For example our LogHost: LogSurfer is scanning the Syslog-Files and if problems are shown by Pattern Matching an information is transferred to NAGIOS.

Some Services are running in an Cluster environment so we have to define how much components may fail before an alarm is triggered, check_cluster Plugin is used.

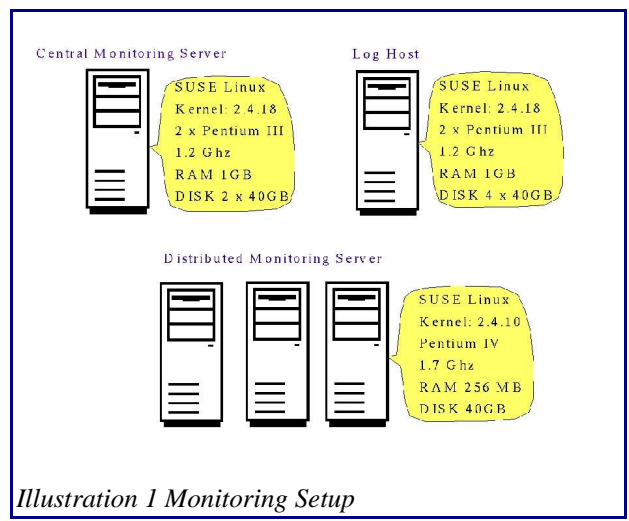

*Illustration 1 Monitoring Setup*

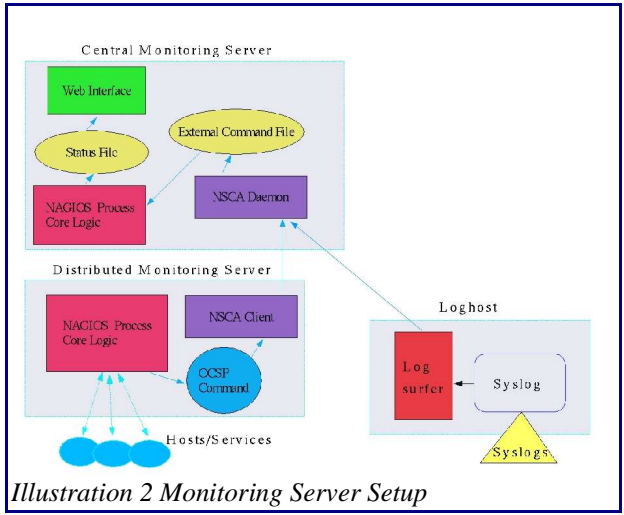
*Illustration 2 Monitoring Server Setup*

NAGIOS stores status and other answers in flat files, to tune up the System we use a RAM-Disk on the Central Monitoring Server, round trip time for all checks is less than one Minute.

Load range is from 0.2 on the Central Monitoring Server up to 3 on the Distributed Server.

### 2.2. Monitoring Policy

Our Monitoring Policy is that every Host in the Computer Center will be monitored and also Centrally Supported Printers. Monitoring for Network Devices will be done by our Network Group, we only check them by Ping.

*Table 1 Monitoring Policy*

| Host | Check by |
|---|---|
| Network Device | PING |
| Farm PC | PING |
| Printer | SNMP |
| Workgroup Server | Load |
| | Disk |
| | Process |
| Mail | POP |
| | IMAP |
| WEB Server | HTTP |
| AFS Server | Service Monitoring |

At present we have 620 Hosts with 1270 Services defined in NAGIOS.

### 2.3. Alarming

Alarms are shown on the Operator Console,, send by Email, SMS or to an Pager.

The open concept of NAGIOS makes it possible to use any Paging or Email Software. For SMS and Pager connection we use SMS-Client.

Table 2 Problem Notification

| |
|---|
| **** Nagios 1.0 ***** |
| Notification Type: PROBLEM Service: IT Web Server |
| Host: WWW Server WEB |
| Address: 131.169.40.38 |
| State: CRITICAL |
| Date/Time: Tue Mar 19 08:35:59 MET 2003 |
| Additional Info: Connection refused by host |

Table 3 Recovery Notification

| |
|---|
| ***** Nagios 1.0 ***** |
| Notification Type: RECOVERY |
| Service: IT Web Server |
| Host: WWW Server WEB |
| Address: 131.169.40.38 |
| State: OK Date/Time: Wed Mar 19 08:37:46 MET 2003 |
| Additional Info: HTTP ok: HTTP/1.1 200 OK - 0 second response time |

### 2.4. Configuration

All our Configuration is done by flat files, so Nagios depends not on any Database.
There are several adons to create the Configurtion Files.
A Calendar Utility was added to specify different alrming hours for the Operator Staff.
A Perl script reads data from our Asset Managent System and create Host and Service Configuration.

*Table 4 Service Test Configuration*

| define service{ | |
|---|---|
| name | fileserver |
| service_description | fileserver |
| is_volatile | 0 |
| active_checks_enabled | 0 |
| passive_checks_enabled | 1 |
| check_period | 24x7 |
| max_check_attempts | 10 |
| normal_check_interval | 1 |
| retry_check_interval | 5 |
| notification_interval | 2200 |
| notification_period | 24x7 |
| notification_options | w,u,c,r |
| check_command | doing some tests |
| register | 0 |
| } | |

**References :**

**NAGIOS**
http://www.nagios.org

**Log Surfer**
http://www.dfn-cert.de/eng/logsurf/index.html

**SMS-Client**
http://www.styx.demon.co.uk/

*Table 5 Host Configuration*

| define hostgroup{ | |
|---|---|
| name | night |
| hostgroup_name | night |
| alias | night |
| contact_groups | sgi-admins |
| members | netra8,test1,test2 |
| } | |
| define host{ | |
| host_name | netra8 |
| alias | netra AFS Server |
| address | 131.169.40.109 |
| parents | route-194,route-40 |
| use | hostcheck |
| } | |